\documentclass[11pt,twoside,onecolumn]{article}
\usepackage[]{amsmath,amssymb}
\usepackage[]{epsfig}
\newcommand{\be}{\begin{eqnarray}}
\newcommand{\ee}{\end{eqnarray}}

\newcommand{\expec}[1]{\mbox{$\langle\, #1\,\rangle$}}

\renewcommand{\d}{\mbox{{\rm d}}}
\title{\bf String Gas Baryogenesis}
\author{G.L.~Alberghi\thanks{e-mail: alberghi@bo.infn.it} $\ $ \\ 
\\
Dipartimento di Fisica Universit\`a di Bologna
\\
I.N.F.N. 
Sezione di Bologna, Italy }
\begin{document}
%
%
\maketitle
\begin{abstract}
%
%
%
%
We describe a possible realization of the spontaneous baryogenesis mechanism
in the context of extra-dimensional string cosmology and specifically in the string gas scenario.

\end{abstract}
%
\pagestyle{plain}
\raggedbottom
\setcounter{page}{1}
%
%
%
%
%
\section{Introduction}
\label{intro}
One of the most peculiar features of the Universe
is the observed baryon asymmetry,  the difference
between the density of baryons and anti-baryons,
quantitatively described by the dimensionless number
\be
{n_B\over s} \equiv \eta
\simeq 10^{-10}
\ ,
\label{eta}
\ee
where $n_B \equiv n_b - n_{\bar b}$ is the difference between the
baryon and anti-baryon density and $s$ is the entropy density.
The primordial nucleosynthesis, which is one of the most consistent and
precise results in the standard model of cosmology, requires this value
for $ \eta $ at the time of light elements
($^3 He$, $^4 He$, $^7 Li$) production and it
is believed not to have changed since.
\par
The necessary conditions for generating the baryonic asymmetry were
formulated by Sacharov in 1967 \cite{Sakharov} (see also
Ref.~\cite{DimopoulosSusskind}) as:
\begin{enumerate}
\item
Different interactions for particles and antiparticles, or, in other words,
a violation of the C and CP symmetries;
\item
Non-conservation of the baryonic charge;
\item
Departure from thermal equilibrium.
\end{enumerate}
The last condition results from an application of the CPT theorem.
In fact, CPT invariance ensures that the energy spectra for baryons and
anti-baryons are identical, leading consequently to an identical
distribution in thermal equilibrium.
This explains why the baryon number asymmetry was required to be generated
out of thermal equilibrium.
\par
The so called spontaneous baryogenesis mechanism \cite{CohenKaplan} uses
the natural CPT non-invariance of the Universe during its early history
to bypass this third condition, thus allowing the baryonic asymmetry to 
be generated in thermal 
equilibrium. We know, in fact,  that an expanding Universe at finite temperature violates both
Lorentz invariance and time reversal, and this can lead to effective CPT
violating interactions \cite{cpt}.
Thus the cosmological expansion of the early Universe leads us naturally
to examine the possibility of generating the baryon asymmetry in thermal
equilibrium.
The main ingredient for this mechanism is a scalar field
$ \phi $ with a derivative coupling to the baryonic current.
If the current is not conserved and the time derivative of the scalar
field  has a non-vanishing expectation value, an effective chemical
potential with opposite signs for baryons and anti-baryons is generated
leading to an asymmetry even in thermal equilibrium.
\par
We applied this mechanism in the context of brane cosmology, by identifying the scalar field
with the radion field  (the size of the extra-dimension) and by using its natural coupling  
to the trace of the four-dimensional stress energy tensor \cite{Alberghi}.
Subsequently this idea was applied in the 
context of four-dimensional  standard cosmology by coupling  the baryonic current
directly to the trace of the stress-energy tensor \cite{Steinhardt}.
\par
Let us now turn to describe the string gas cosmology framework:
string gas cosmology is an approach towards studying the 
effects of superstring theory on early universe 
cosmology and is based on new symmetries and new degrees of freedom of string theory. 
Within this context, it appears possible to stabilize the moduli which describe the size and shape of the
 extra spatial dimensions without the need of introducing many extra tools such as warping and fluxes. 
In fact, in the absence of a non-perturbative formulation of string theory,
the approach to string cosmology 
is to focus on symmetries and degrees of freedom which are new to
string theory (compared to point particle theories) and which will
be part of any non-perturbative string theory, and to use
them to develop a new cosmology. 
In particular the symmetry is T-duality and the new degrees of freedom are the string
oscillatory modes and the string winding modes.

String gas cosmology is based on coupling a classical background
which includes the graviton and the dilaton fields to a gas of
strings (and possibly other basic degrees of freedom of
string theory such as ``branes"). 
If for simplicity, one  takes all spatial directions to be toroidal and
 denote the radius of the torus by $R$, strings have three types
of states:  momentum modes which represent the center
of mass motion of the string, oscillatory modes which
represent the fluctuations of the strings, and winding
modes counting the number of times a string wraps the torus.
The second key feature of string theory upon which string gas cosmology
is based is T-duality. To introduce this symmetry, let us discuss the
radius dependence of the energy of the basic string states:
The energy of an oscillatory mode is independent of $R$, momentum
mode energies are quantized in units of $1/R$, i.e.
$ E_n \, = \, n  /R $   
and winding mode energies are quantized in units of $R$, i.e.
$ E_m \, = \, m R  $
where both $n$ and $m$ are integers. Thus, a new symmetry of
the spectrum of string states emerges: Under the change $ R \, \rightarrow \, 1/R $
in the radius of the torus (in units of the string length $l_s$)
the energy spectrum of string states is
invariant if winding and momentum quantum numbers are interchanged
$ (n, m) \, \rightarrow \, (m, n)  $.
The above symmetry is the simplest element of a larger
symmetry group, the T-duality symmetry group which in
general also mixes fluxes and geometry. Postulating
that T-duality extends to non-perturbative string theory leads
to the need of adding D-branes to the list of fundamental
objects in string theory. With this addition, T-duality is expected
to be a symmetry of non-perturbative string theory.

Given the outlined framework of String Cosmology,
the question we would like to address in this article is whether the mechanism
of spontanous baryogenesis might be at work in such models.

In Section 2 we describe the spontaneous baryogenesis mechanism, in Section 3
we show the basic equations of String Gas Cosmology and in Section 4
we show how  the spontaneous baryogenesis mechanism might effectively be at work in the String Gas Cosmology scenario.

%
%
\section{Spontaneous Baryogenesis}
\label{SPsection}
To illustrate the mechanism of spontaneous baryogenesis
(see e.g.~Refs.~\cite{CohenKaplan,Trodden,china}) let us consider
a theory in which a neutral scalar field $\phi$ is coupled to the baryonic
current $ J^{\mu}_B $ by the Lagrangian density
\be
L_{\rm int} = {\lambda' \over M_c }\, J^{\mu}_B\,\partial _{\mu} \phi
\ ,
\ee
where $\lambda '$ is a coupling constant and $M_c <M_{\rm Pl} $ is a
cut-off mass scale in the theory.
Let us assume that $ \phi $ is homogeneous, so that only the time
derivative term contributes,
\be
L_{\rm int} = { \lambda ' \over M_c }\,\dot \phi\, n_B
\equiv \mu(t)\,  n_B
\ ,
\label{Lint}
\ee
where $ n_B = J^0_B $ is the baryon number density and $ \mu(t) $ is to
be regarded as an effective time-dependent chemical potential.
This interpretation (see Ref.~\cite{Dolgov}) is valid if the current
$ J^{\mu}_B $ is not conserved (otherwise one could integrate the
interaction term away) and if $ \phi $ behaves as an external field which
develops a slowly varying time derivative $ \expec{\dot \phi} \neq 0 $
as the Universe expands.
Since the chemical potential $\mu$ enters with opposite signs for baryons
and anti-baryons, we have a net baryonic charge density in thermal equilibrium
at the temperature $T$,
\be
 n_B (T; \xi ) = \int { \d^3 k \over (2 \pi)^3 }\,
\left[ f(k,\mu) - f(k,-\mu) \right]
\ ,
\ee
where $ \xi \equiv \mu / T $ is regarded as a parameter, and
\be
f(k, \mu) =
{ 1 \over \exp \left[ \left( \sqrt{k^2 + m^2} - \mu \right) /T \right]
\pm 1}
\label{distrib}   
\ee
is the phase-space thermal distribution~\footnote{The plus sign is for
fermions and the minus sign for bosons.}
for particles with rest mass $m$ and momentum $ k $.
For $|\xi|\ll 1 $ we may expand Eq.~(\ref{distrib}) in powers of
$ \xi $  to obtain
\be
n_B (T; \mu)={g\,T^3 \over 6 }\,\xi + O\left( \xi ^2\right)
\ ,
\ee
where $g$ is the number of degrees of freedom of the field corresponding
to $n_B$.
Upon substituting in for the expression of $\mu$, one therefore finds
\be
n_B (T;\mu) \simeq { \lambda'\, g \over 6\,M_c }\, T^2\,
\expec{\dot\phi}
\ .
\label{asymmetry}
\ee
\par
Whatever the mechanism of baryon number violation, we assume there is a
temperature $T_F$ at which the baryon number violating processes become
sufficiently rare so that $n_B$ freezes out (we will call $T_F$ the
{\em freezing temperature\/}).
Once this temperature is reached as the universe cools down, one is left
with a baryon asymmetry whose value is given by Eq.~(\ref{asymmetry})
evaluated at $T=T_F$.
The value of the parameter $ \eta $ remains unchanged in the subsequent
evolution.
%
%

\section{String Gas Cosmology}

The equations of String Gas Cosmology are based on coupling  an ideal gas of
string and brane modes, described by an energy density $\rho$
and pressures $p_i$ in the i'th spatial direction,  to the background space-time of dilaton gravity. 
They follow from the 10-dimensional string frame action
\be
\label{dilgrav}
S \, = \, {1 \over {2 \kappa^2}} \int d^{10}x \sqrt{-g} e^{-2 \phi}
\bigl[{\hat R} + 4\, \partial^{\mu} \phi \, \partial_{\mu} \phi \bigr] + S_m \, ,
\ee
where $g$ is the determinant of the metric, ${\hat R}$ is the Ricci scalar,
$\phi$ is the dilaton, 
$\kappa$ is the reduced gravitational constant in ten dimensions,
and $S_m$ denotes the matter action. 
Let us thus first consider radion stabilization in the string frame
\cite{Watson2}. For this purpose, the ansatz for the metric is
\be
ds^2 \, = \, dt^2 - e^{2 \lambda} d{\bf x}^2 - e^{2 \nu} d{\bf y}^2 \, ,
\ee
where ${\bf x}$ are the coordinates of the three large dimensions and
${\bf y}$ the coordinates of the internal dimensions.
The variational equations of motion which follow from the dilaton
gravity action (\ref{dilgrav}) for the above  metric are
\be
- 3 {\ddot \lambda} - 3 {\dot \lambda}^2 - 6 {\ddot \nu} - 6 {\dot \nu}^2
+ 2 {\ddot \phi} \, &=& \, {1 \over 2} e^{2 \phi} \rho \\
{\ddot \lambda} + 3 {\dot \lambda}^2 + 6 {\dot \lambda} {\dot \nu}
- 2 {\dot \lambda} {\dot \phi} \, &=& \, {1 \over 2} e^{2 \phi} p_{\lambda} \\
{\ddot \nu} + 6 {\dot \nu}^2 + 3 {\dot \lambda} {\dot \nu}
- 2 {\dot \nu} {\dot \phi} \, &=& \, {1 \over 2} e^{2 \phi} p_{\nu} \\
- 4 {\ddot \phi} + 4{\dot \phi}^2 - 12 {\dot \lambda}{\dot \phi}
- 24 {\dot \nu} {\dot \lambda} + 3 {\ddot \lambda} + 6 {\dot \lambda}^2
+ 6 {\ddot \nu} + 21 {\dot \nu}^2 + 18 {\dot \lambda} {\dot \nu} \, &=& 
\, 0 
\ee
where $\rho$ is the energy density and $p_{\lambda}$ and $p_{\nu}$
are the pressure densities in the non-compact and compact directions, 
respectively.

Let us consider, for simplicity, a superposition of several string gases, one with
momentum number $M_3$ about the three large dimensions, one with
momentum number $M_6$ about the six internal dimensions, and a further
one with winding number $N_6$ about the
internal dimensions. 
the energy $E$ and the total pressures $P_{\lambda}$ and $P_{\nu}$ are given by
\be
E \, &=& \, 
\mu \bigl[ 3  M_3 e^{- \lambda} + 6 M_6 e^{- \nu} + 6 N_6 e^{\nu} \bigr] \\
P_{\lambda} \, &=& \, \mu M_3 e^{- \lambda} \\ 
P_{\nu} \, &=& \, \mu \bigl[ - n_6 e^{\nu} + M_6 e^{- \nu} \bigr] \, ,
\label{EA3}
\ee
where $\mu$ is the string mass per unit length. One might consider
a more realistic string gas made up of strings which have momentum, winding and oscillatory
quantum numbers together. The states considered here are massive,
and would not be expected to dominate the thermodynamical partition
function if there are states which are massless. However, for the
purpose of studying radion stabilization in the string frame, the
use of the above naive string gas is sufficient.

In the symmetric case $M_6 = N_6$
it follows from (\ref{EA3}) that the equation of motion for
$\nu$ is a damped oscillator equation, with the minimum of the effective
potential corresponding to the self-dual radius. The damping is due
to the expansion of the three large dimensions, driven by the pressure
from the momentum modes $N_3$. We thus see that the
naive intuition that the competition of winding and momentum modes about
the compact directions stabilizes the radion degrees of freedom at the
self-dual radius generalizes to this anisotropic setting.

\subsection{Radion Stabilization in the Einstein Frame}

In order to make contact with observational cosmology, it is important to
consider the issue of radion stabilization when the dilaton is frozen, 
or, more generally, in the Einstein frame. As discussed in 
\cite{Patil1,Patil2} (see also earlier comments in \cite{Watson2}),
the existence of string modes which are massless at the self-dual radius
is crucial in obtaining radion stabilization in the Einstein frame.

Let us then consider the equations of motion which arise from coupling
the Einstein action to a string gas. In the anisotropic setting  the metric
is taken to be
\be
ds^2 \, = \, dt^2 - a(t)^2 d{\bf x}^2 - 
\sum_{\alpha = 1}^6 b_{\alpha}(t)^2 dy_{\alpha}^2 \, ,
\ee
where the $y_{\alpha}$ are the internal coordinates and the
equation of motion for $b_{\alpha}$ becomes
\be \label{extra}
{\ddot b_{\alpha}} + 
\bigl( 3 H + \sum_{\beta = 1, \beta \neq \alpha}^6 {{{\dot b_{\beta}}} \over {b_{\beta}}} \bigr) {\dot b_{\alpha}} \, = \, 
\sum_{n, m} 8 \pi G {{\mu_{m,n}} \over {\sqrt{g} \epsilon_{m,n}}}{\cal S} \,
\ee
where $\mu_{m,n}$ is the number density of $(m,n)$ strings, $\epsilon_{m,n}$
is the energy of an individual $(m,n)$ string, and $g$ is the determinant of
the metric. The source term ${\cal S}$ depends on the quantum numbers of the
string gas, and the sum runs over all momentum numbers and winding
number vectors $m$ and $n$, respectively (note that $n$ and $m$ are
six-vectors, one component for each internal dimension). 

Let us take the special case where all $ b_ {\alpha} = b $ and write the metric as
\be
ds^2 \, = \, dt^2 - a(t)^2 d{\bf x}^2 - b(t)^2 d{\bf y}^2 \, ,
\ee 

We will analyse the evolution in a four-dimensional
effective field theory, where we replace the radion field $b(t)$ by a scalar
field $\varphi(t)$. In order that $\varphi$ be canonically normalized
when starting from the higher dimensional action of General Relativity,
$\varphi$ and $b$ must be related as (\cite{WatBatrev})
\be
\varphi \, = \, M_{p} \, \sqrt{2d} \,  \log  b 
\ee
where $M_{p}$ is the four-dimensional Planck mass. If the bulk
size starts out at the string scale, then $b(t_i) = 1$, where $t_i$
is the initial time. With these normalizations, $\varphi = 0$
corresponds to string separation between the branes. 
\par
The next crucial step is to invoke a mechanism to
stabilize the radius of the extra dimensions at a fixed radius. Such
modulus stabilization mechanisms have recently been extensively studied both
in the context of string theory models of inflation (see e.g. 
\cite{stringinflation} for recent reviews)
and in string gas cosmology \cite{RHBrev3}. We will make use of the
mechanism developed in the latter approach.

String modes which carry momentum about the extra dimensions will generate
an effective potential for the radion which is repulsive. These
repulsive effects will dominate for values of the radion smaller than
the self-dual radius. Since these modes are very light at large values of
the radion, it is likely that they will be present in great abundance.
Even if they are not, the subset of such modes which are massless at
enhanced symmetry points will be copiously produced when the value
of the radion approaches such points \cite{Watson,Beauty}.

The induced potential will lead to a source term in the equation
of motion for the scale factor $b(t)$ which is of the form 
\cite{Patil1,Patil2}
\be
{\ddot b} + 3H {\dot b} \, = \, 8 \pi G n(t) 
\bigl[\bigl({1 \over b}\bigr)^2 - b^2 \bigr] + ... \, ,
\ee
where the dots indicate extra source terms from other string modes,
as well as terms quadratic in ${\dot b}$. Note that $n(t)$ is given
by the number density of the modes. Translating to the
scalar field $\varphi$, and neglecting terms quadratic in ${\dot \varphi}$,
the above equation becomes
\be
{\ddot \varphi} + 3H {\dot \varphi} \, = \, 8 \pi G n(t)
e^{-\varphi / M_{p}} M_{p} 
\bigl( e^{-2\varphi / M_{p}} - e^{2\varphi / M_{p}} \bigr) \, .
\ee
Thus, it follows that after approaching the self-dual radius, $b(t)$
will perform damped oscillations about $b(t) = 1$, or, in other words,
$\varphi(t)$ will undergo damped oscillations about and get
trapped at $\varphi = 0$ (which corresponds to string scale separation
between the orbifold fixed planes).
Interestingly this is the same kind of motion obtained in the string frame.
 
%
%
%
%

\section{Baryogenesis}

Following the discussion on \cite{Alberghi, Trodden} let us now assume that the high energy Lagrangian 
contains an interaction term of the form of Eq.~(\ref{Lint}),
\be
L_{\rm int} = { \lambda \over  M_c }\, \dot  \phi \, n_B
\ ,
\ee
where (see \cite{Natalia2})
\be 
\phi = \sqrt{6 } \,  M_p \log{ b }
\ee
is the canonically normalized scalar field describing the size of the extra-dimensions.
From the results of Section 2 one has  
\be
   \eta = {n_B \over s} = {\lambda \over 6 \, M_c} {g \over g_*} { \expec{\dot\phi}  \over T} 
\ee
Let us now consider the stabilization of the extra-dimension in this context:
neglecting small corrections, the evolution of $ \sigma (t) $ is that of damped harmonic oscillator
as previously described, with damping given by the Hubble parameter 
\be
      { \dot b  \over b }  = \expec {\dot  \phi }  \simeq 3 H
\ee  
One can straightforwardly obtain
an expression for the baryonic asymmetry at temperature T
\be
   \eta \simeq  \lambda\;  { g   \; M_p  \over  g_*  \;  M_c } \; {H \over T} 
\ee
with $ H ( T ) $ evaluated at the freezing temperature $T _ F $.
In a phase of radiation domination the request that this value be compatible with the
observed baryonic asymmetry, would imply $ T > 10^{-10 } M_c $ .
On the other hand for a gas of strings one has (see  \cite {Easson})
\be
    H(T) \simeq  c_{s } \, T
\ee
where $c_{s}$ is a constant. This implies that the ratio $ H / T $ is  a constant and the 
constraint for this mechanism to be compatible with the experimental data is translated in a constraint
on the constant $ c_{s} $
\be
   c_{s} \,  M_p \ge 10^{-10} M_c
\ee
Finally in the case of brane gas the Hubble parameter is proportional to the square root of the
temperature
\be
   H(T) \simeq  c_b\,  \sqrt{T}
\ee
where $c_b$ is a constant. One then has a constraint
\be
    T_F  \le 10 ^{20} \left(  c_b \,  {M_p \over M_c}  \right) ^2 
\ee 
This case is peculiar in the sense that it gives an upper bound on the freezing temperature
whereas one usually obtains lower bounds.


%
%
%

%
%

\section{Conclusions}
We have shown that the evolution and stabilization of extra-dimensions
within the framework of string gas cosmology provides a natural scenario for the 
realization of the spontaneous baryogenesis mechanism.
This allows us to extract  bounds on the temperature at which
the baryonic asymmetry might be generated.
\bigskip
\label{conc}
\par
\par
\noindent

%
%
%
%

%
\end{document}